\documentclass[a4paper]{llncs}

\usepackage{multicol}
\usepackage[english]{babel}

\usepackage{color}
\usepackage{xcolor}

\usepackage{amsmath}
\usepackage{amssymb}
\usepackage{caption}
\usepackage{subcaption}
\RequirePackage[noline,noend,linesnumbered]{algorithm2e}

\usepackage{multirow}

\newcommand\dint[3]{\Delta_{#1}^{#2,#3}}

\include{mathdefs}

\auxfun{prepare}
\auxfun{effect}
\auxfun{eval}
\auxfun{apply}
\auxfun{merge}
\auxfun{inc}
\auxfun{dec}
\auxfun{rd}
\auxfun{add}
\auxfun{rmv}
\auxfun{elements}

\usepackage{xspace}
\newcommand\dsrdt{\hbox{$\delta$-CRDT}\xspace}

\auxfun{size}

\usepackage{soul}

\begin{document}


\title{Delta State Replicated Data Types}
%
\author{Paulo S\'{e}rgio Almeida \and Ali Shoker\and Carlos Baquero}
\institute{HASLab/INESC TEC and Universidade do Minho, Portugal}

\maketitle

\begin{abstract}

CRDTs are distributed data types that make eventual consistency of a
distributed object possible and non ad-hoc. Specifically, state-based CRDTs
ensure convergence through disseminating the entire state, that may be large,
and merging it to other replicas; whereas operation-based CRDTs disseminate
operations (i.e., small states) assuming an exactly-once reliable
dissemination layer.
We introduce \emph{Delta State Conflict-Free Replicated Data Types} (\dsrdt)
that can achieve the best of both worlds: small messages with an incremental
nature, as in operation-based CRDTs, disseminated over unreliable
communication channels, as in traditional state-based CRDTs.  This is achieved
by defining \emph{$\delta$-mutators} to return a \emph{delta-state}, typically
with a much smaller size than the full state, that to be joined with both
local and remote states.
We introduce the \dsrdt framework, and we explain it through establishing
a correspondence to current state-based CRDTs.
In addition, we present an anti-entropy algorithm for eventual convergence,
and another one that ensures causal consistency. Finally, we introduce several
\dsrdt specifications of both well-known replicated datatypes and novel datatypes,
including a generic map composition.

\end{abstract}

\section{Introduction}
\label{sec:intro}

Eventual consistency (EC) is a relaxed consistency model that is often adopted
by large-scale distributed
systems~\cite{riak+crdt,syn:optim:rep:1433,app:rep:optim:1606} where
availability must be maintained, despite outages and partitioning, whereas
delayed consistency is acceptable. 
A typical approach in EC systems is to allow replicas of a distributed object
to temporarily diverge, provided that they can eventually be reconciled into a
common state. To avoid application-specific reconciliation methods, costly and
error-prone, \emph{Conflict-Free Replicated Data
Types}~(CRDTs)~\cite{rep:syn:sh138,syn:rep:sh143} were introduced, allowing the
design of self-contained distributed data types that are always available and
eventually converge when all operations are reflected at all replicas. Though
CRDTs are deployed in practice and support millions of users worldwide~\cite{soundcloud,lol,bet365}, more work is
still required to improve their design and performance.


CRDTs support two complementary designs: \emph{operation-based} (or op-based) and \emph{state-based}.
In op-based designs~\cite{alg:rep:sh132,syn:rep:sh143}, the execution of an operation is
done in two phases: \emph{prepare} and \emph{effect}. The former is
performed only on the local replica and looks at the operation and current
state to produce a message that aims to represent the operation, which is then
shipped to all replicas.  Once received, the representation of the
operation is applied remotely using \emph{effect}. 
On the other hand, in a state-based
design~\cite{app:1639,syn:rep:sh143} an operation is only executed on the
local replica state. A replica periodically propagates its local changes to
other replicas through shipping its entire state. A received state is
incorporated with the local state via a \emph{merge} function that
deterministically reconciles both states. To maintain convergence,
\emph{merge} is defined as a \emph{join}: a least upper bound over a
join-semilattice~\cite{app:1639,syn:rep:sh143}.

Op-based CRDTs have some advantages as they can allow for simpler 
implementations, concise replica state, and smaller messages; however, 
they are subject to some limitations: First, they assume a message dissemination
layer that guarantees reliable exactly-once causal broadcast; these guarantees are hard to
maintain since large logs must be retained to prevent duplication even if TCP 
is used~\cite{idempotenceHelland:2012}.
Second, 
membership management is a hard task in op-based systems especially once the number 
of nodes gets larger or due to churn problems, since 
all nodes must be coordinated by the middleware. Third, the op-based approach requires 
operations to be executed individually (even when batched) on all nodes.


The alternative is to use state-based systems, which are free from these 
limitations. 
However, a major drawback in current state-based CRDTs is the communication overhead of
shipping the entire state, which can get very large in size. For instance, the
state size of a \emph{counter} CRDT (a vector of integer counters, one per
replica) increases with the number of replicas; whereas in a \emph{grow-only
Set}, the state size depends on the set size, that grows as more operations are
invoked.
This communication overhead limits the use of state-based CRDTs to data-types
with small state size (e.g., counters are reasonable while large sets are not).
Recently, there has been a demand for CRDTs with large state sizes (e.g., in
RIAK DT Maps~\cite{Brown:2014:RDM:2596631.2596633} that can compose multiple CRDTs and that we formalize in Section \ref{sec:map}).

In this paper, we rethink the way state-based CRDTs should be designed, having
in mind the problematic shipping of the entire state. Our aim is to ship a
\emph{representation of the effect} of recent update operations on the state,
rather than the whole state, while preserving the idempotent nature of
\emph{join}.  This ensures convergence over unreliable communication (on the
contrary to op-based CRDTs that demand exactly-once delivery and are prone to
message duplication).  To achieve this, we develop in detail the concept of
\emph{Delta State-based CRDTs} (\dsrdt) that we initially introduced in
\cite{DBLP:conf/netys/AlmeidaSB15}. In this new (delta) framework, the state is
still a join-semilattice that now results from the join of multiple
fine-grained states, i.e., \emph{deltas}, generated by what we call
\emph{$\delta$-mutators}. \emph{$\delta$-mutators} are new versions of the
datatype mutators that return the effect of these mutators on the state. In
this way, deltas can be temporarily retained in a buffer to be shipped
individually (or joined in groups) instead of shipping the entire object. The
changes to the local state are then incorporated at other replicas by joining
the shipped deltas with their own states. 

The use of ``deltas'' (i.e., incremental states) may look intuitive in state
dissemination; however, this is not the case for state-based CRDTs.  The reason
is that once a node receives an entire state, merging it locally is simple
since there is no need to care about causality, as both states are
self-contained (including meta-data).  The challenge in \dsrdt is that
individual deltas are now ``state fragments'' and usually must be causally
merged to maintain the desired semantics.  This raises the following questions:
is merging deltas semantically equivalent to merging entire states in CRDTs? If
not, what are the sufficient conditions to make this true in general? And under
what constraints causal consistency is maintained?  This paper answers these
questions and presents corresponding proofs and examples.


We address the challenge of designing a new \dsrdt that conserves the
correctness properties and semantics of an existing CRDT by establishing a
relation between the novel $\delta$-mutators with the original CRDT mutators.
We prove that eventual consistency is guaranteed in \dsrdt as long as all
deltas produced by $\delta$-mutators are delivered and joined at other
replicas, and we present a corresponding simple anti-entropy algorithm.  We
then show how to ensure causal consistency using deltas through introducing
the concept of \emph{delta-interval} and the \emph{causal delta-merging
condition}. Based on these, we then present an anti-entropy algorithm for
\dsrdt, where sending and then joining delta-intervals into another replica
state produces the same effect as if the entire state had been shipped and
joined. 

We illustrate our approach through a simple $counter$ CRDT and a corresponding $\dsrdt$
specification. Later, we present a portfolio of several ${\dsrdt}s$ that adapt known CRDT designs and also introduce a generic kernel for the definition of CRDTs that keep a causal history of known events and a CRDT map that can compose them. All these $\dsrdt$ datatypes, and a few more, are available online in a reference C++ library~\cite{deltaCode}. 
Our experience shows that a $\dsrdt$ version can be devised for all CRDTs,
but this requires some design effort that varies with the complexity of
different CRDTs. This refactoring effort can be avoided for new datatypes by
writing all mutations as delta-mutations, and only deriving the standard
mutators if needed; these can be trivially obtained from the delta-mutators.

\section{System Model}

Consider a distributed system with nodes containing local memory, with no
shared memory between them.
Any node can send messages to any other node. The network is asynchronous;
there is no global clock, no bound on the time a message takes to
arrive, and no bounds on relative processing speeds. The network is unreliable:
messages can be lost, duplicated or reordered (but are not corrupted). Some
messages will, however, eventually get through: if a node sends infinitely
many messages to another node, infinitely many of these will be delivered. In
particular, this means that there can be arbitrarily long partitions, but
these will eventually heal.
Nodes have access to durable storage; they can crash but will eventually 
recover with the content of the durable storage just before the crash occurred.
Durable state is written atomically at each state transition.
Each node has access to its globally unique identifier in a set $\ids$.

\subsection{Notation}

We use mostly standard notation for sets and maps, including set
comprehension of the forms $\{f(x) | x \in S\}$ or $\{x \in S | Pred(x)\}$.
A map is a set of $(k, v)$ pairs, where each $k$ is associated with a single
$v$. Given a map $m$, $m(k)$ returns the value associated with key $k$,
while $m\{k \mapsto v\}$ denotes $m$ updated by mapping $k$ to $v$.
The domain and range of a map $m$ is denoted by $\dom m$ and $\ran m$,
respectively, i.e., $\dom m = \{k | (k,v) \in m\}$ and $\ran m = \{v | (k,v)
\in m\}$.
We use $\fst p$ and $\snd p$ to denote the first and second component of a
pair $p$, respectively.
We use $\bool$, $\nat$, and $\integer$, for the booleans, natural numbers, and
integers, respectively;
also $\ids$ for some unspecified set of node identifiers.
Most sets we use are partially ordered and have a least element $\bot$ (the
bottom element).
We use $A \map B$ for a partial function from $A$ to $B$; given such a map
$m$, then $\dom m \subseteq A$ and $\ran m \subseteq B$, and for convenience
we use $m(k)$ when $k \not\in \dom m$ and $B$ has a bottom, to denote
$\bot_B$; e.g., for some $m : \ids \map \nat$, then $m(k)$ denotes $0$ for any
unmapped key $k$.



%
%
%
%
%
%

\section{A Background of State-based CRDTs}
\label{sec:back}
\emph{Conflict-Free Replicated Data Types}~\cite{rep:syn:sh138,syn:rep:sh143} (CRDTs) are distributed datatypes that allow different replicas of a distributed CRDT instance to diverge and ensures that, eventually, all replicas converge to the same state. State-based CRDTs achieve this through propagating updates of the local state by disseminating the entire state across replicas. The received states are then merged to remote states, leading to convergence (i.e., consistent states on all replicas).

A state-based CRDT consists of a triple $(S, M, Q)$, where $S$ is a
join-semi\-lattice~\cite{lattices:Brian:2002}, $Q$ is a set of query functions (which return some result without modifying the state), and $M$ is a set of mutators that perform updates; a mutator $m \in M$ takes a state $X \in S$ as input and returns a new state $X' = m(X)$. A join-semilattice is a set with a \emph{partial
order} $\pleq$ and a binary \emph{join} operation $\join$ that returns the
\emph{least upper bound }(LUB) of two elements in $S$; a \emph{join} is designed to be commutative, associative, and idempotent. Mutators are defined in such a way to be \emph{inflations}, i.e., for any mutator $m$ and state $X$, the following holds:
\[ X \pleq m(X) \]
In this way, for each replica there is a monotonic sequence of states, defined under the lattice partial order, where each subsequent state subsumes the previous state when joined elsewhere.

Both query and mutator operations are always available since they are performed
using the local state without requiring inter-replica communication; however,
as mutators are concurrently applied at distinct replicas, replica states will
likely diverge. Eventual convergence is then obtained using an
\emph{anti-entropy} protocol that periodically ships the entire local state to
other replicas.  Each replica merges the received state with its local state
using the \emph{join} operation in $S$.  Given the mathematical properties of
\emph{join}, if mutations stop being issued and anti-entropy proceeds, all replicas eventually converge to
the same state. i.e. the least upper-bound of all states involved.  State-based
CRDTs are interesting as they demand little guarantees from the dissemination
layer, working under message loss, duplication, reordering, and temporary
network partitioning, without impacting availability and eventual convergence.

Fig.~\ref{fig:statectr} represents a state-based increment-only counter. The
$\sf{GCounter}$ CRDT state is a map from replica identifiers to positive integers.
Initially, the bottom state $\bot$ is an empty map (unmapped keys implicitly
mapping to zero). A single mutator, i.e., $\inc_i$, is defined that increments
the value corresponding to the local replica $i$ (returning the updated map).
The query operation $\af{value}$ returns the counter value by adding the
integers in the map entries.  The join of two states is the point-wise maximum
of the maps.
Mutators, like $\inc_i$, are in general parameterized by the replica id $i$, so
that their exact behavior can depend on it, while queries, like $\af{value}$,
are typically replica agnostic and only depend on the CRDT state, regardless
of in which replica they are invoked.

\begin{figure}[t]
\begin{eqnarray*}
  \sf{GCounter} & = & \ids \map \nat \\
\bot & = & \{ \} \\
\inc_i(m) & = &  m\{i \mapsto m(i)+1\}\\
\af{value}(m) & = & \sum_{j \in \ids} m(j) \\
m \join m' & = & \{ j \mapsto \max(m(j),m'(j)) | j \in \dom m \union \dom m' \}
\end{eqnarray*}
\caption{State-based Counter CRDT; replica $i$.}
\label{fig:statectr}
\end{figure}

The main weakness of state-based CRDTs is the cost of dissemination of updates,
as the full state is sent. In this simple example of counters, even though
increments only update the value corresponding to the local replica $i$, the
whole map will always be sent in messages, even when the other map entries
remained unchanged (e.g., if no messages have been received and merged).

It would be interesting to only ship the recent modification incurred on the
state, and possibly any received modifications that effectively changed it. This is, however, not possible with the current model of state-based
CRDTs as mutators always return a full state. Approaches which simply ship
operations (e.g., an ``increment $n$'' message), like in operation-based
CRDTs, require reliable communication (e.g., because increment is not
idempotent).
In contrast, the modification that we introduce in the next section allows producing and encoding recent mutations in an
incremental way, while keeping the advantages of the state-based approach,
namely the idempotent, associative, and commutative properties of join.

\section{Delta-state CRDTs}
\label{sec:model}


We introduce \emph{Delta-State Conflict-Free Replicated Data Types}, or \dsrdt
for short, as a new kind of state-based CRDTs, in which \emph{delta-mutators}
are defined to return a \emph{delta-state}: a value in the same
join-semilattice which represents the updates induced by the mutator on the
current state.


\begin{definition}[Delta-mutator]
A delta-mutator $m^\delta$ is a function, corresponding to an update operation, which takes a state $X$ in a
join-semilattice $S$  as parameter and returns a delta-mutation $m^\delta(X)$, also in $S$.
\end{definition}

\begin{definition}[Delta-group]
A delta-group is inductively defined as either a delta-mutation or a join of
several delta-groups.
\end{definition}

\begin{definition}[\dsrdt]
A \dsrdt consists of a triple $(S, M^\delta, Q)$, where $S$ is a
join-semilattice, $M^\delta$ is a set of delta-mutators, and $Q$ a set of
query functions, where the state transition at each replica is given by either
joining the current state $X \in S$ with a delta-mutation:

\[
X' = X \join m^\delta(X),
\]
or joining the current state with some received delta-group $D$:
\[
X' = X \join D.
\]
\end{definition}


In a \dsrdt, the effect of applying a mutation, represented by a
delta-mutation $\delta = m^\delta(X)$, is decoupled from the resulting state
$X' = X \join \delta$, which allows shipping this $\delta$ rather than the
entire resulting state $X'$.  All state transitions in a $\dsrdt$, even upon
applying mutations locally, are the result of some join with the current state.
Unlike standard CRDT mutators, delta-mutators do not need to be inflations in order to inflate a state; this is however ensured by joining their output, i.e., deltas, into the current state: $X \pleq X \join m^\delta(X)$.




In principle, a delta could be shipped immediately to remote replicas once
applied locally.
For efficiency reasons, multiple deltas returned by applying several
delta-mutators can be joined locally into a delta-group and retained in
a buffer. The delta-group can then be shipped to remote replicas to be joined
with their local states. Received delta-groups can optionally be joined into
their buffered delta-group, allowing transitive propagation of
deltas. A full state can be seen as a special (extreme) case of a delta-group.



If the causal order of operations is not important and the intended aim is
merely eventual convergence of states, then delta-groups can be shipped using
an unreliable dissemination layer that may drop, reorder, or duplicate
messages.  Delta-groups can always be re-transmitted and re-joined, possibly
out of order, or can simply be subsumed by a less frequent sending of the full
state, e.g., for performance reasons or when doing state transfers to new
members.

\subsection{Delta-state decomposition of standard CRDTs}

A \dsrdt $(S,M^\delta,Q)$ is a \emph{delta-state decomposition} of a 
state-based CRDT $(S,M,Q)$, if for every mutator $m \in M$, we have a
corresponding mutator $m^\delta \in M^\delta$ such that, for every state $X \in
S$:

\[
m(X) = X \join m^\delta(X)
\]

This equation states that applying a delta-mutator and joining into the
current state should produce the same state transition as applying the
corresponding mutator of the standard CRDT.

Given an existing state-based CRDT (which is always a trivial decomposition of
itself, i.e., $m(X) = X \join m(X)$, as mutators are inflations), it will be
useful to find a non-trivial decomposition  such that delta-states returned
by delta-mutators in $M^\delta$ are smaller than the resulting 
state: \[ \size(m^\delta(X)) \ll \size(m(X)) \]

In general, there are several possible delta-state decompositions, with
multiple possible delta-mutators that correspond to each standard
mutator. In order to minimize the generated delta-states (which will typically
minimize their size) each delta-mutator chosen $m^\delta$ should be minimal
in following sense: for any other alternative choice of delta-mutator
$m^{\delta'}$,
for any $X$, $m^{\delta'}(X) \not\ple  m^\delta(X)$.
Intuitively, minimal delta-mutators do not leak into the deltas they produce
any redundant information that is already present in $X$.
Moreover (although in theory not always necessarily the case) for typical
datatypes that we have come across in practice, for each mutator $m$ there
exists a corresponding \emph{minimum} delta-mutator $m^{\delta_\bot}$, i.e.,
with $m^{\delta_\bot} \pleq m^{\delta'}$ (under the standard pointwise
function comparison), for any alternative delta-mutator. As we will see in the
concrete examples, typically minimum delta-mutators are found naturally, with
no need for some special ``search''.




\subsection{Example: \dsrdt Counter}
\label{sec:counter}

Fig.~\ref{fig:deltactr} depicts a \dsrdt
specification of a counter datatype that is a delta-state decomposition of the
state-based counter in Fig.~\ref{fig:statectr}.  The state, join and
$\af{value}$ query operation remain as before.  Only the mutator $\inc^\delta$
is newly defined, which increments the map entry corresponding to the local
replica and only returns that entry, instead of the full map as $\inc$ in the
state-based CRDT counter does.  This maintains the original semantics of the
counter while allowing the smaller deltas returned by the delta-mutator to be
sent, instead of the full map.  As before, the received payload (whether one or
more deltas) might not include entries for all keys in $\ids$, which are assumed
to have zero values.  The decomposition is easy to understand in this example since the equation $\inc_i(X) = X \join \inc_i^\delta(X)$ holds as $m\{i \mapsto m(i)+1\}= m \join \{i \mapsto m(i)+1\}$. In other words, the
single value for key $i$ in the delta, corresponding to the local replica
identifier, will overwrite the corresponding one in $m$ since the former maps to a higher value (i.e., using $\max$).
Here it can be noticed that: (1) a delta \emph{is} just a state, that can be joined possibly several times without requiring exactly-once delivery, and without being a
representation of the ``increment'' operation {(as in operation-based CRDTs), which is itself non-idempotent; (2) joining deltas into a delta-group and disseminating delta-groups at a lower rate than the operation rate reduces data communication overhead, since multiple increments from a given source can be collapsed into a single state counter. 

  \begin{figure}[t]
\begin{eqnarray*}
  \af{GCounter} & = & \ids \map \nat \\
  \bot & = & \{ \} \\
  \inc_i^\delta(m) & = & \{i \mapsto m(i)+1\} \\
  \af{value}(m) & = & \sum_{j \in \ids} m(j) \\
  m \join m' & = & \{ j \mapsto \max(m(j),m'(j)) | j \in \dom m \union \dom m' \}
\end{eqnarray*}
\caption{A \dsrdt counter; replica $i$.} 
\label{fig:deltactr}
\end{figure}

\section{State Convergence}
\label{sec:conv}



In the \dsrdt execution model, and regardless of the anti-entropy
algorithm used, a replica state always evolves by joining the current
state with some \emph{delta}: either the result of a delta-mutation, or some
arbitrary delta-group (which itself can be expressed as a join of
delta-mutations). Without loss of generality, we assume $S$ has a bottom
$\bot$ which is also the initial state. (Otherwise, a bottom can always be
added, together with a special $\af{init}$ delta-mutator, which returns the
initial state.) Therefore, all states can be expressed as joins of
delta-mutations, which makes state convergence in \dsrdt easy to achieve: it
is enough that all delta-mutations generated in the system reach every
replica, as expressed by the following proposition.

\begin{proposition}(\dsrdt convergence)
\label{prop:convergence}
Consider a set of replicas of a \dsrdt object, replica $i$ evolving along a
sequence of states $X_i^0=\bot, X_i^1, \ldots$, each replica performing
delta-mutations of the form $m_{i,k}^\delta(X_i^k)$ at some subset of its
sequence of states, and evolving by joining the current state either
with self-generated deltas or with delta-groups received from others. If each
delta-mutation $m_{i,k}^\delta(X_i^k)$ produced at each replica is joined
(directly or as part of a delta-group) at least once with every
other replica, all replica states become equal.
\end{proposition}

\begin{proof}
Trivial, given the associativity, commutativity, and idempotence of the join
operation in any join-semilattice.
\end{proof}

This opens up the possibility of having anti-entropy algorithms that are only devoted to enforce convergence, without necessarily providing causal consistency (enforced in standard CRDTs); thus, making a trade-off between performance and consistency guarantees.  For
instance, in a counter (e.g., for the number of \emph{likes} on a social
network), it may not be critical to have causal consistency, but merely not to
lose increments and achieve convergence.


\subsection{Basic Anti-Entropy Algorithm}
A basic anti-entropy algorithm that ensures eventual convergence in \dsrdt is
presented in Algorithm~\ref{alg:basic-algo}. For the node corresponding to
replica $i$, the durable state, which persists after a crash, is simply the
\dsrdt state $X_i$. The volatile state $D$ stores a delta-group that is used
to accumulate deltas before eventually sending it to other replicas.
The initial value for both $X_i$ and $D_i$ is $\bot$.



{
\auxfun{send}
\auxfun{receive}
\auxfun{operation}
\auxfun{ack}
\auxfun{random}

\begin{algorithm}[t]

\begin{multicols}{2}
\DontPrintSemicolon
\SetKwBlock{inputs}{inputs:}{}
\SetKwBlock{dstate}{durable state:}{}
\SetKwBlock{vstate}{volatile state:}{}
\SetKwBlock{periodically}{periodically}{}
\SetKwBlock{on}{on}{}

\inputs{
  $n_i \in \pow\ids$, set of neighbors \;
  $t_i \in \bool$, transitive mode\;
  $\af{choose}_i \in S \times S \to S$, state/delta\;
}

\dstate{
  $X_i \in S$, CRDT state, $X_i^0 = \bot$ \;
}

\vstate{
  $D_i \in S$, delta-group, $D_i^0 = \bot$ \;
}

\on({$\operation_i(m^\delta)$}){
  $d = m^\delta(X_i)$ \;
  $X_i' = X_i \join d$ \;
  $D_i' = D_i \join d$ \;
}

\BlankLine
\on({$\receive_{j,i}(d)$}){
  $X_i' = X_i \join d$ \;
  \uIf{$t_i$}{ $D_i' = D_i \join d$ \; }
  \uElse { $D_i' = D_i$ \; }
}

\BlankLine
\periodically(){
  $m = \af{choose}_i(X_i, D_i)$ \;
  \For{$j \in n_i$}{
    $\send_{i,j}(m)$ \;
  }
  $D_i' = \bot$ \;
}

\end{multicols}
\bigskip
\caption{Basic anti-entropy algorithm for \dsrdt. \label{alg:basic-algo}}
\end{algorithm}
}

When an operation is performed, the corresponding delta-mutator $m^\delta$ is
applied to the current state of $X_i$, generating a delta $d$. This delta is joined
both with $X_i$ to produce a new state, and with $D$.
In the same spirit of standard state based CRDTs, a node sends its messages in a periodic fashion, where the message payload is either the delta-group $D_i$ or the full state $X_i$; this decision is made by the function $\af{choose}_i$ which returns one of them. To keep the algorithm simple, a node simply broadcasts its messages without distinguishing between neighbors. After each send, the delta-group is reset to $\bot$.

Once a message is received, the payload $d$ is joined into the current \dsrdt
state.  The basic algorithm operates in one of two modes: (1) a
\emph{transitive} mode (when $t_i$ is true) in which $d$ is also joined into
$D$, allowing transitive propagation of delta-mutations, where deltas
received at node $i$ from some node $j$ can later be sent to some other node
$k$; (2) a \emph{direct} mode where a delta-group is exclusively the join of
local delta-mutations ($j$ must send its deltas directly to $k$). The
decisions of whether to send a delta-group versus the full state (typically
less periodically), and whether to use the transitive or direct mode are out
of the scope of this paper. In general, decisions can be made considering many
criteria like delta-group size, state size, message loss distribution
assumptions, and network topology.

\section{Causal Consistency}
\label{sec:causal}


Traditional state-based CRDTs converge using joins of the full state, which
implicitly ensures per-object causal consistency~\cite{DBLP:conf/popl/BurckhardtGYZ14}: each state of
some replica of an object reflects the causal past of operations on the object
(either applied locally, or applied at other replicas and transitively
joined).

Therefore, it is desirable to have \dsrdt{}s offer the same causal-consistency guarantees that standard state-based CRDTs offer. This raises the question about how can delta propagation and merging of \dsrdt be constrained (and expressed in an anti-entropy algorithm) in such a manner to give the same results as if a standard state-based CRDT was used. Towards this objective, it is useful to define a particular kind of delta-group, which we call a \emph{delta-interval}:

 \begin{definition}[Delta-interval]
Given a replica $i$ progressing along the states
$X_i^0, X_ i^1 , \ldots$, by joining delta $d_i^k$ (either local
delta-mutation or received delta-group) into $X_i^k$ to obtain
$X_i^{k+1}$, a delta-interval $\dint iab$ is a delta-group resulting from
joining deltas $d_i^a, \ldots, d_i^{b-1}$:
 \[
\dint iab = \bigjoin \{ d_i^k | a \leq k < b \}
 \]
 \end{definition}

The use of delta-intervals in anti-entropy algorithms will be a key ingredient
towards achieving causal consistency. We now define a restricted kind of
anti-entropy algorithm for \dsrdt{}s.

\begin{definition}[Delta-interval-based anti-entropy algorithm]
A given anti-entropy algorithm for \dsrdt{}s is delta-interval-based, if all
deltas sent to other replicas are
delta-intervals.
\end{definition}

Moreover, to achieve causal consistency the next condition must satisfied:

\begin{definition}[Causal delta-merging condition]
\label{def:c_cond}
A delta-interval based anti-entropy algorithm is said to satisfy the causal
delta-merging condition if the algorithm only joins $\dint jab$ from replica
$j$ into state $X_i$ of replica $i$ that satisfy:
\[
X_i \pgeq X_j^a.
\]
\end{definition}


This means that a delta-interval is only joined into states that at least
reflect (i.e., subsume) the state into which the first delta in the interval
was previously joined. The causal delta-merging condition is important, since
any delta-interval based anti-entropy algorithm for a \dsrdt that satisfies it
can be used to obtain the same outcome of a standard CRDT; this is formally
stated in Proposition~\ref{prop:corr}.

\begin{proposition}(CRDT and \dsrdt correspondence)
\label{prop:corr}
Let $(S,M,Q)$ be a standard state-based CRDT and $(S,M^\delta,Q)$ a corresponding delta-state
decomposition. Any \dsrdt state reachable by an execution $E^\delta$
over $(S,M^\delta,Q)$, by a delta-interval based anti-entropy algorithm $A^\delta$
satisfying the causal delta-merging condition, is equal to a state
resulting from an execution $E$ over $(S,M,Q)$, having the corresponding data-type
operations, by an anti-entropy algorithm $A$ for state-based CRDTs.
\end{proposition}

\begin{proof}

By simulation, establishing a correspondence between an execution $E^\delta$,
and execution $E$ of a standard CRDT of which $(S,M^\delta,Q)$ is a decomposition,
as follows:
1) the state $(X_i, D_i, \ldots)$ of each node in $E^\delta$ containing
CRDT state $X_i$, information about delta-intervals $D_i$ and possibly other
information, corresponds to only $X_i$ component (in the same join-semilattice);
2) for each action which is a delta-mutation $m^\delta$ in $E^\delta$, $E$ executes
he corresponding mutation $m$, satisfying $m(X) = X \join m^\delta(X)$;
3) whenever $E^\delta$ contains a send action of a delta-interval $\dint iab$,
execution $E$ contains a send action containing the full state $X_i^b$;
4) whenever $E^\delta$ performs a join into some $X_i$ of a delta-interval
$\dint jab$, execution $E$ delivers and joins the corresponding message
containing the full CRDT state $X_j^b$.
By induction on the length of the trace, assume that for each replica $i$,
each node state $X_i$ in $E$ is equal to the corresponding component in the
node state in $E^\delta$, up to the last action in the global trace.
A send action does not change replica state, preserving the correspondence.
Replica states only change either by performing data-type update operations or
upon message delivery by merging deltas/states respectively. If the next
action is an update operation, the correspondence is preserved
due to the delta-state decomposition property $m(X) = X \join m^\delta(X)$.
If the next action is a message delivery at replica $i$, with a merging
of delta-interval/state from other replica $j$, because algorithm $A^\delta$
satisfies the causal merging-condition, it only joins into state $X_i^k$ a
delta-interval $\dint jab$ if $X_i^k \pgeq X_j^a$. In this case, the outcome
will be:
\begin{eqnarray*}
X_i^{k+1}
     &=& X_i^k \join \dint jab \\
     &=& X_i^k \join \bigjoin \{ d_j^l | a \leq l < b \}\\
     &=& X_i^k \join X_j^a \join \bigjoin \{ d_j^l | a \leq l < b \}\\
     &=& X_i^k \join X_j^a \join d_j^a \join d_j^{a+1} \join \ldots \join d_j^{b-1} \\
     &=& X_i^k \join X_j^{a+1} \join d_j^{a+1} \join \ldots \join d_j^{b-1} \\
     &=& \ldots \\
     &=& X_i^k \join X_j^{b-1} \join d_j^{b-1} \\
     &=& X_i^k \join X_j^b \\
\end{eqnarray*}
The resulting state $X_i^{k+1}$ in $E^\delta$ will be, therefore, the same as
the corresponding one in $E$ where the full CRDT state from $j$ has been joined,
preserving the correspondence between $E^\delta$ and $E$.

\end{proof}


\begin{corollary}(\dsrdt causal consistency)
\label{corol:causal-consistency}
Any \dsrdt in which states are propagated and joined using a delta-interval-based anti-entropy algorithm satisfying the causal delta-merging condition ensures causal consistency.
\end{corollary}

\begin{proof}
From Proposition~\ref{prop:corr} and causal consistency of state-based CRDTs. 
\end{proof}

\subsection{Anti-Entropy Algorithm for Causal Consistency}
Algorithm~\ref{alg:causal-algo} is a delta-interval based anti-entropy
algorithm which enforces the causal delta-merging condition. It can be used
whenever the causal consistency guarantees of standard state-based CRDTs are
needed. For simplicity, it excludes some optimizations that are
important in practice, but easy to derive. The algorithm distinguishes
neighbor nodes, and only sends to each one appropriate delta-intervals that
obey the delta-merging condition and can joined at the receiving node.


{
\auxfun{send}
\auxfun{receive}
\auxfun{operation}
\auxfun{ack}
\auxfun{random}


\begin{algorithm}[t]
\begin{multicols}{2}
\DontPrintSemicolon
\SetKwBlock{inputs}{inputs:}{}
\SetKwBlock{dstate}{durable state:}{}
\SetKwBlock{vstate}{volatile state:}{}
\SetKwBlock{periodically}{periodically}{}
\SetKwBlock{on}{on}{}

\inputs{
  $n_i \in \pow\ids$, set of neighbors \;
}

\dstate{
  $X_i \in S$, CRDT state, $X_i^0 = \bot$ \;
  $c_i \in \nat$, sequence number, $c_i^0 = 0$ \;
}

\vstate{
  $D_i \in \nat \map S$, deltas, $D_i^0 = \{\}$ \;
  $A_i \in \ids \map \nat$, ack map, $A_i^0 = \{\}$ \;
}

\on({$\receive_{j,i}(\af{delta}, d, n)$}){
  \uIf{$d \not \pleq X_i$}{
    $X_i' = X_i \join d$ \;
    $D_i' = D_i \{ c_i \mapsto d \}$ \;
    $c_i' = c_i + 1$ \;
  }
  $\send_{i,j}(\ack, n)$ \;
}

\on({$\receive_{j,i}(\ack, n)$}){
  $A_i' = A_i \{ j \mapsto \max(A_i(j), n)\}$ \;
}

\on({$\operation_i(m^\delta)$}){
  $d = m^\delta(X_i)$ \;
  $X_i' = X_i \join d$ \;
  $D_i' = D_i \{ c_i \mapsto d \}$ \;
  $c_i' = c_i + 1$ \;
}

\periodically(// ship interval or state){
  $j = \random(n_i)$ \;
  \uIf{$D_i = \{\} \lor \min \dom D_i > A_i(j)$}{
    $d = X_i$
  } \uElse{
    $d = \bigjoin\{ D_i(l) | A_i(j) \leq l < c_i \}$ \;
  }
  \uIf{$A_i(j) < c_i$}{
    $\send_{i,j}(\af{delta}, d, c_i)$ \;
  }
}

\periodically(// garbage collect deltas){
  $l = \min \{ n | (\_, n) \in A_i \}$ \;
  $D_i' = \{ (n,d) \in D_i | n \geq l \}$ \;
}
\BlankLine
\end{multicols}
\bigskip
\caption{Anti-entropy algorithm ensuring causal consistency of \dsrdt.}
\label{alg:causal-algo}

\end{algorithm}
}

Each node $i$ keeps a contiguous sequence of deltas $d_i^l, \ldots, d_i^u$ in a
map $D$ from integers to deltas, with $l = \min \dom D$ and $u =
\max \dom D$. The sequence numbers of deltas are obtained from the counter
$c_i$ that is incremented when a delta (whether a delta-mutation or
delta-interval received) is joined with the current state.  Each node $i$
keeps an acknowledgments map $A$ that stores, for each neighbor $j$, the
largest index $b$ for all delta-intervals $\dint iab$ acknowledged by $j$
(after $j$ receives $\dint iab$ from $i$ and joins it into $X_j$).

Node $i$ sends a delta-interval $d=\dint iab$ with a $(\af{delta}, d,
b)$ message; the receiving node $j$, after joining $\dint iab$ into its
replica state, replies with an acknowledgment message $(\af{ack}, b)$; if an
ack from $j$ was successfully received by node $i$, it updates the entry of
$j$ in the acknowledgment map, using the $\max$ function. This handles
possible old duplicates and messages arriving out of order.

Like the \dsrdt state, the counter $c_i$ is also kept in a durable storage.
This is essential to cope with potential crash and recovery incidents.
Otherwise, there would be the danger of receiving some delayed ack, for a
delta-interval sent before crashing, causing the node to skip sending some
deltas generated after recovery, thus violating the delta-merging condition.

The algorithm for node $i$ periodically picks a random neighbor $j$. In
principle, $i$ sends the join of all deltas starting from the latest delta
acked by $j$. Exceptionally, $i$ sends the entire state in two
cases: (1) if the sequence of deltas $D_i$ is empty, or (2) if $j$ is
expecting from $i$ a delta that was already removed from $D_i$ (e.g., after a
crash and recovery, when both deltas and the ack map, being volatile state,
are lost). A delta message is only sent if the counter $c_i$ has advanced
past the next delta expected by node $j$, i.e., if $A_i(j) < c_i$, to avoid sending
the full state in local inactivity periods, when no local operations are being
issued, all neighbor nodes have acked all deltas, and garbage collection has
been applied, making the $D_i$ map empty. To garbage collect old deltas, the
algorithm periodically removes the deltas that have been acked by \emph{all}
neighbors.


\begin{proposition}
Algorithm~\ref{alg:causal-algo} produces the same reachable states as a
standard algorithm over a CRDT for which the \dsrdt is a decomposition,
ensuring causal consistency.
\label{prop:causal-algo}
\end{proposition}

\begin{proof}
From Proposition~\ref{prop:corr} and Corollary~\ref{corol:causal-consistency},
it is enough to prove that the algorithm satisfies the causal delta-merging
condition. The algorithm explicitly keeps deltas $d_i^k$ tagged with
increasing sequence numbers (even after a crash), according with the
definition; node $j$ only sends to $i$ a delta-interval $\dint jab$ if $i$ has
acked $a$; this ack is sent only if $i$ has already joined some delta-interval
(possibly a full state) $\dint jka$. Either $k = 0$ or, by the same reasoning,
this $\dint jka$ could only have been joined at $i$ if some other interval
$\dint jlk$ had already been joined into $i$. This reasoning can be recursed
until a delta-interval starting from zero is reached. Therefore, $X_i \pgeq
\bigjoin \{ d_j^k | 0 \leq k < a \} = \dint j0a = X_j^a$.

\end{proof}

\newcommand\auxfunpar[1]{\expandafter\newcommand\csname #1\endcsname[1]{%
 \mathop{\hbox{$\mathsf{#1}$}}\nolimits\langle ##1 \rangle}}

\newcommand\afpar[2]{\af{#1}\langle #2 \rangle}

\section{Portfolio of \dsrdt{}s}
\label{sec:portfolio}


Having established the equivalence to classic state based CRDTs we now derive
a series of specifications based on delta-mutators. Although we cover a
significant number of CRDTs, the goal is not to provide an exhaustive survey,
but instead to illustrate more extensively the design of specifications with
deltas. In our experience the intellectual effort of designing a delta-based
CRDT is not much higher than designing it with standard mutators. Since
standard mutators can be obtained from delta-mutators, by composing these with
join, having delta-mutators as basic building blocks can only add flexibility
to the designs.

First, we will cover simple CRDTs and CRDT compositions that do not require
distinguished node identifiers for the mutation. Next, we cover CRDTs that
require a unique identifier for each replica that is allowed to mutate the
state, and make use of this identifier in one or more of the mutations. Then,
we address the important class of what we denote by \emph{Causal CRDTs},
presenting a generic design in which the state lattice is formed by a
\emph{dot store} and a \emph{causal context}. We define three such dot stores
and corresponding lattices, which are then used to defined several causal CRDTs.
We conclude the portfolio with a Map design, a causal CRDT which can correctly
embed any causal CRDT, including the map itself.

All of the selected CRDTs have delta implementations available in
C++~\cite{deltaCode}, that complement the specifications. Most of the Causal
CRDTs, including the Map, are also available in Erlang and deployed in
production as part of \textsf{Riak DT}~\cite{riakDT}.

\subsection{Simple Lattice Compositions}

To obtain composite CRDTs, a basic ingredient is being able to obtain states,
which are join-semilattices, as composition of join-semilattices. Two common
useful cases are the product and lexicographic product. Other examples of
lattice composition are presented in \cite{Kemme:2014:DSR:2596583.2596601,lattices:Brian:2002}.

\subsubsection{Pair}
In Figure \ref{fig:Pair} we show the standard pair composition.
The bottom is the pair of respective bottoms and the join is the
coordinate-wise join of the components. This can be generalized to products of
more than two components.

\begin{figure}[t]
\begin{eqnarray*}
  \af{Pair}\langle A,B \rangle & = & A \times B  \\ 
  \bot & = & (\bot,\bot) \\
  (a,b) \join (a',b') & = & (a \join a', b \join b' )
\end{eqnarray*}
\caption{Pair of join-semilattices.}
\label{fig:Pair}
\end{figure}

\subsubsection{Lexicographic Pair} A variation of the \emph{pair} composition
is to establish a \emph{lexicographic pair}. In this construction, in Figure
\ref{fig:LexPair}, the first element takes priority in establishing the
outcome of the join, and a join of the second component is only performed
on a tie. One important special case in when the first component is a total
order; it can be used, e.g., to define an outcome based on the comparison of a
time-stamp, as will be shown later.

\begin{figure}[t]
\begin{eqnarray*}
  \af{LexPair} \langle A,B \rangle  & = & A \boxtimes B \\
  \bot & = & (\bot,\bot) \\
  (a,b) \join (a',b') & = &
  \begin{cases}
   (a,b)        & \text{if } a > a' \\
   (a',b')      & \text{if } a' > a \\
   (a, b \join b')        & \text{if } a = a' \\
   (a \join a', \bot)        & \text{otherwise}
  \end{cases}
\end{eqnarray*}
\caption{Lexicographic pair of join-semilattices.}
\label{fig:LexPair}
\end{figure}

\subsection{Anonymous \dsrdt{}s}

The simplest CRDTs are anonymous. This occurs when the mutators do not make
use of a globally unique replica identifier, having a uniform specification
for all replicas. (Although for uniformity of notation we will keep
parameterizing mutators by replica identifier.)

\subsubsection{GSet}
A simple example is illustrated by a grow-only set, in Figure \ref{fig:GSet}.
The single delta mutator $\af{insert}_i^\delta(e,s)$ does not even need to
consider the current state of the replica, available in $s$, and simply
produces a delta with a singleton set containing the element $e$ to be added.
This delta $\{e\}$ when joined to $s$ produces the desired result: an inflated
set $s \cup \{e\}$ that includes element $e$. The \emph{join} of grow-only
sets is trivially obtained by unioning the sets.

\begin{figure}[t]
\begin{eqnarray*}
  \af{GSet}\langle E \rangle & = & \pow{E}\\
\bot & = & \{\} \\
  \af{insert}_i^\delta(e,s) & = & \{e\} \\
\elements(s) & = & s\\
s \join s' & = & s \cup s'
\end{eqnarray*}
\caption{\dsrdt grow-only set, replica $i$.}
\label{fig:GSet}
\end{figure}

\subsubsection{2PSet} In case one needs to remove elements, there are multiple
ways of addressing it. The simplest way is to include another (grow-only) set
that gathers the removed elements. This is done in Figure \ref{fig:2PSet},
which shows a \emph{two-phase set}, with state being a pair of sets.
The name comes from the fact that elements may go through two phases: the
\emph{added} phase and the \emph{removed} phase. The shortcoming of this
simple design is that once removed, elements cannot be re-added.


\begin{figure}[t]
\begin{eqnarray*}
  \af{2PSet}\langle E \rangle & = & \pow{E} \times \pow{E} \\
\bot & = & (\bot,\bot) \\
  \af{insert}_i^\delta(e,(s,t)) & = & (\{e\},\bot) \\
  \af{remove}_i^\delta(e,(s,t)) & = & (\bot, \{e\})\\
  \elements((s,t)) & = & s \setminus t\\
  (s,t) \join (s',t') & = & (s \join s', t \join t')
\end{eqnarray*}
\caption{\dsrdt two-phase set, replica $i$.}
\label{fig:2PSet}
\end{figure}

If we look at the query function $\elements$ it is clear
that the data-type is presenting to the user the set difference between the
added elements and the removed elements (those stored in the tombstone set
$t$). Removing an element simply adds it to the \emph{removed} set.
(A variant of 2PSet with \emph{guarded removes} \cite{DBLP:conf/forte/ZellerBP14} only does so if the element is already present in the \emph{added}
set.) The join is simply a pairwise join.

\subsubsection{Add-Wins Last-Writer-Wins Set} This construction, depicted in
Figure \ref{fig:AWLWWSet}, manages a set of elements of type $E$ by tagging
them with timestamps from some total order -- here we use natural numbers.
Each time an elements is added, it is tagged with a client supplied timestamp
and the boolean $\true$. Removed elements are similarly tagged, but with
the boolean $\false$. Elements marked with $\true$ are considered to be
in the set. When joining two such sets, those elements in common will have to
compete to define if they are in the set.
By using lexicographic pairs, we obtain the behaviour that elements with higher
(more recent) time-stamps will win, defining the presence according to the
boolean tag; if there is a tie in the time-stamp, adds will win, since we
order $\false < \true$.

\begin{figure}[t]
\begin{eqnarray*}
  \af{AWLWWSet}\langle E \rangle & = & E \map \nat \boxtimes \bool \\
  \bot & = & \{ \} \\
  \af{insert}_i^\delta(e,t,m) & = & \{ e \mapsto (t,\true)\} \\
  \af{remove}_i^\delta(e,t,m) & = & \{ e \mapsto (t,\false)\}\\
  \elements(m) & = & \{ e \mid (e, (\_,\true)) \in m\}\\
  m \join m' & = & \{ e \mapsto m(e) \join m'(e) \mid e \in \dom m \union \dom m'\}
\end{eqnarray*}
\caption{\dsrdt Add-Wins LWW Set, replica $i$.}
\label{fig:AWLWWSet}
\end{figure}

Notice that is is up to the client to ensure that supplied timestamps always grow monotonically. Failure to do so is a common source of errors in timestamp based systems \cite{doomstone}. A dual construction to the \emph{Add-Wins LWW Set} is a \emph{Remove-Wins LWW Set}, where remove operations take priority on the event of a time-stamp tie. This construction has been widely deployed in production as part the SoundCloud system \cite{soundcloud}. 

\subsection{Named \dsrdt{}s}

Another design strategy for conflict-free data-types is to ensure that each
replica only changes a specific part of the state. In Section \ref{sec:model},
we defined a \textsf{GCounter} that, using a map from globally unique replica
identifiers to natural numbers, keeps track of how many increment operations
each replica did. This was the first example of a \emph{named CRDT}, the
construction covered in this section. The distinction from anonymous CRDTs is
that mutations make use of the replica identifier $i$.

\subsubsection{PNCounter} By composing, in a pair, two grow-only counters we
obtain a \emph{positive-negative counter} that can track both increments and
decrements. Shown in Figure \ref{fig:PNCounter}, the increment and decrement
operations will update the first and second components of the pair,
respectively. As expected, the value is obtained by subtracting the decrements
from
the increments.

\begin{figure}[t]
\begin{eqnarray*}
  \af{PNCounter} & = & \af{GCounter} \times \af{GCounter}\\
  \bot & = & (\bot, \bot) \\
  \inc_i^\delta((p,n)) & = & (\inc_i^\delta(p),\bot) \\
  \dec_i^\delta((p,n)) & = & (\bot,\inc_i^\delta(n)) \\
  \af{value}((p,n)) & = & \af{value}(p) - \af{value}(n)\\
  (p,n) \join (p',n') & = & ( p \join p', n \join n' )
\end{eqnarray*}
\caption{\dsrdt positive-negative counter, replica $i$.}
\label{fig:PNCounter}
\end{figure}

\subsubsection{Lexicographic Counter} While the \textsf{PNCounter} was one of
the first CRDTs to be added to a production database, in Riak
1.4~\cite{riak14}, the competing Cassandra database had its own counter
implementations based on the LWW strategy. Interestingly it proved to be
difficult to avoid semantic anomalies in the behaviour of those early counters,
and since Cassandra 2.1, a new counter was
introduced~\cite{cassandracounters}.  We capture its main properties in the
Figure \ref{fig:LexCounter} specification of a \textsf{LexCounter}.

\begin{figure}[t]
\begin{eqnarray*}
  \af{LexCounter} & = & \ids \map \nat \boxtimes \integer \\
  \bot & = & \{\} \\
  \inc_i^\delta(m) & = & \{i \mapsto m(i)+(0,1)\} \\
  \dec_i^\delta(m) & = & \{i \mapsto m(i)+(1,-1)\} \\
  \af{value}(m) & = & \sum_{j \in \ids} \snd  m(j)  \\
  m \join m' & = & \{ j \mapsto m(j) \join m'(j)  | j \in \dom m \union \dom m' \}
\end{eqnarray*}
\caption{\dsrdt Lexicographic Counter, replica $i$.}
\label{fig:LexCounter}
\end{figure}

This counter is updated by either incrementing or decrementing the second
component of the lexicographic pair corresponding to the replica issuing the
mutation. Decrements also increment the first component, to ensure
that the pair will be inflated, making it (and therefore, the just updated
second component) win upon a lexicographic join.


%
%
%
%

\auxfun{CausalContext}
\auxfun{DotStore}
\auxfun{DotSet}
\auxfun{Lattice}
\auxfun{next}
\auxfunpar{DotMap}
\auxfunpar{DotFun}
\auxfunpar{Causal}


\subsection{Causal \dsrdt{}s}

We now introduce a specific class of CRDTs, that we will refer to as
\emph{causal CRDTs}. Initial designs \cite{syn:rep:sh143} introduced data
types such as \emph{observed-remove sets} and \emph{multi-value registers}.
While these made possible sets which allow adding and removing elements
multiple times, and to model the design of the eventually consistent shopping
cart, in Amazon Dynamo \cite{app:rep:optim:1606}, they had sub-optimal
scalability properties \cite{DBLP:conf/popl/BurckhardtGYZ14}.  Later designs,
such as in \emph{observed-remove sets without tombstones}
\cite{rep:opt:sh151}, allow an efficient management of meta-data state and can
be applied to a broad class of data-types.

We introduce the concept of \emph{dot store} to be used together with a
\emph{causal context} to form the state (a join-semilattice) of a causal CRDT,
presenting three such dot stores and lattices. These are then used to obtain
several related data-types, including flags, registers, sets, and maps.

\subsubsection{Causal Context}

A common property to causal CRDTs is that events can be assigned unique
identifiers. A simple mechanism is to create these identifiers
by appending to a globally unique replica identifier a replica-unique integer.
For instance, in replica $i \in \ids$ we can create the sequence $(i,1),
(i,2), \dots$. Each of these pairs can be used to tag a specific event, or
client action, and if we collect these pairs in a grow-only set, we can
remember which events are known to each replica. The pair is called a
\emph{dot} and the grow-only set of pairs can be called a \emph{causal
history}, or alternatively a \emph{causal context}, as we do here.

As seen in Figure~\ref{fig:CausalContext}, a causal context is a set of dots.
We define two functions over causal contexts: $\max_i(c)$ gives the maximum
sequence number for pairs in $c$ from replica $i$, or $0$ if there is no such
dot; $\af{next}_i(c)$ produces the next available sequence number for replica
$i$ given set of events in $c$.

\begin{figure}[t]
\begin{eqnarray*}
  \CausalContext &=& \pow{\ids \times \nat} \\
  \max_i(c) & = &  \max(\{n | (i,n) \in c\} \cup \{0\})\\
  \next_i(c) &=& (i, \max_i(c) + 1)
\end{eqnarray*}
\caption{Causal Context.}
\label{fig:CausalContext}
\end{figure}

\subsubsection{Causal Context Compression}

In practice, a causal context can be efficiently compressed without any loss
of information. When using an anti-entropy algorithm that provides causal
consistency, e.g., Algorithm~\ref{alg:causal-algo}, then for each replica
state $X_i$ that includes a causal context $c_i$, and for any replica
identifier $j \in \ids$, we have a contiguous sequence:
\[
1 \leq n \leq \max_j(c_i) \implies (j,n) \in c_i.
\]
Thus, under causal consistency the causal context can always be encoded as a
compact \emph{version vector}~\cite{Parker:1983:DMI:1313337.1313753} $\ids \map \nat$
that keeps the maximum sequence number for each replica.

Even under non-causal anti-entropy, such as in Algorithm \ref{alg:basic-algo},
compression is still possible by keeping a version vector that encodes the
initial contiguous sequence of dots from each replica, together with a set for
the non-contiguous dots. As anti-entropy proceeds, each dot is eventually
encoded in the vector, and thus the set remains typically small. Compression
is less likely for the causal context of delta-groups in transit or buffered
to be sent, but those contexts are only transient and smaller than those in
the actual replica states. Moreover, the same techniques that encodes
contiguous sequences of dots can also be used for transient context
compression~\cite{MGS14}.

\subsubsection{Dot Stores}

Together with a causal context, the state of a causal CRDT will use some kind
of dot store, which acts as a container for data-type specific information.
A dot store can be queried about the set of event identifiers (dots)
corresponding to the relevant operations in the container, by function
$\af{dots}$, which takes a dot store and returns a set of dots. In
Figure~\ref{fig:DotStore} we define three kinds of dot stores: a $\DotSet$ is
simply a set of dots; the generic $\DotFun{V}$ is a map from dots to some
lattice $V$; the generic $\DotMap{K,V}$ is a map from some set $K$ into some
dot store $V$.

\begin{figure}[t]
\begin{eqnarray*}
  \DotStore \\
  \afpar{dots}{S : \DotStore} &:& S \to \pow{\ids \times \nat} \\
\\
  \DotSet : \DotStore &=& \pow{\ids \times \nat} \\
  \af{dots}(s) &=& s \\
\\
  \DotFun{V : \Lattice} : \DotStore &=& \ids \times \nat \map V \\
  \af{dots}(s) &=& \dom s \\
\\
  \DotMap{K, \; V : \DotStore} : \DotStore &=& K \map V \\
  \af{dots}(m) &=& \bigunion\{ \af{dots}(v) | (\_,v) \in m\} \\
\end{eqnarray*}
\caption{Dot Stores.}
\label{fig:DotStore}
\end{figure}

\subsubsection{Causal \dsrdt{}s}

In figure~\ref{fig:DotStoreLattices} we define the join-semilattice which
serves as state for Causal \dsrdt{}s, where an element is a pair of dot store
and causal context. We define the join operation for each of the three kinds
of dot stores.
These lattices are a generalization of techniques introduced
in~\cite{rep:opt:sh151,DBLP:conf/dais/AlmeidaBGPF14}.
To understand the meaning of a state (and the way join must behave), a dot
present in a causal context but not in the corresponding dot store, means that
the dot was present in the dot store, some time the past, but has been removed
meanwhile. Therefore, the causal context can track operations with remove
semantics, while avoiding the need for individual tombstones.

\begin{figure}[t]
\begin{eqnarray*}
  \Causal{T : \DotStore} &=& T \times \CausalContext \\
  \\
  \join &:& \Causal{T} \times \Causal{T} \to \Causal{T} \\
  \\
  \kw{when} && T : \DotSet \\
  (s,c) \join (s',c') &=& ((s \cap s') \union (s \setminus c') \union (s' \setminus c), c \union c') \\
  \\
  \kw{when} && T : \DotFun{\_} \\
  (m,c) \join (m',c')
    &=& (\{ k \mapsto m(k) \join m'(k) | k \in \dom m \cap \dom m' \} \union \\
    && \hphantom( \{ (d,v) \in m | d \not\in c' \} \union
                  \{ (d,v) \in m' | d \not\in c \} , c \union c') \\
  \\
  \kw{when} && T : \DotMap{\_,\_} \\
  (m,c) \join (m',c') &=& (\{ k \mapsto \af{v}(k) | k \in \dom m \union \dom
m' \land \af{v}(k) \neq \bot \}, c \union c')\\
                      && \kw{where} \af{v}(k) = \fst{((m(k),c) \join (m'(k),c'))}
\end{eqnarray*}
\caption{Lattice for Causal \dsrdt{}s.}
\label{fig:DotStoreLattices}
\end{figure}

When joining two replicas, a dot present in only one dot store, but included
in the causal context of the other, will be discarded.
This is clear for the simpler case of a $\DotSet$, where the join preserves
all dots in common, together with those not present in the other causal context.
The $\DotFun{V}$ case is analogous, but the container is now a map from dots to
some value, allowing the value for a given dot to evolve with time,
independently at each replica. It assumes the value set is a join-semilattice,
and applies the corresponding join of values for each dot in common.

In the more complex case of $\DotMap{K,V}$, a map from some $K$ to some
dot store $V$, the join, for each key present in either replica, performs a
join in the lattice $\Causal{V}$, by pairing the per-key value
with the replica-wide causal context, and storing the resulting value (first
component of the result) for that key, but only when it is not $\bot_V$.
This allows the disassociation of a composite embedded value from a key, with
no need for a per-key tombstone, by remembering in the causal context all dots
from the composite value. Matching our notation, in a $\DotMap{K,V}$, any
unmapped key corresponds effectively to the bottom $\bot_V$.

\subsubsection{Enable-Wins Flag} The flags are simple, yet useful, data-types
that were first introduced in Riak 2.0 \cite{riakDT}. Figure \ref{fig:EWFlag}
presents an \emph{enable-wins flag}. Enabling the flag simply replaces all
dots in the store by a new dot; this is achieved by obtaining the dot through
$\next_i(c)$, and making the delta mutator return a store containing just the new
dot, together with a causal context containing both the new dot and all
current dots in the store; this will make all current dots to be removed from
the store upon a join (as previously defined), while the new dot is added.
Concurrent enabling can lead to the store containing several dots. Reads will
consider the flag enabled if the store is not an empty set.
Disabling is similar to enabling, in that all current dots are removed from
the store, but no new dot is added. It is possible to construct a dual
data-type with \emph{disable-wins} semantics and its code is also available
\cite{deltaCode}.

\begin{figure}[t]
\begin{eqnarray*}
  \af{EWFlag}  &=& \Causal{\DotSet} \\
  \af{enable}_i^\delta((s,c)) &=& (d, d \union s)
  \quad \kw{where} d = \{\af{next}_i(c)\} \\
  \af{disable}_i^\delta((s,c)) &=& (\{\}, s) \\
  \af{read}((s,c)) &=& s \neq \{\}
\end{eqnarray*}
\caption{\dsrdt Enable-wins Flag, replica $i$.}
\label{fig:EWFlag}
\end{figure}

\subsubsection{Multi-Value Register} A \emph{multi-value register} supports
read and write operations, with traditional sequential semantics.  Under
concurrent writes, a join makes a subsequent read return all concurrently
written values, and a subsequent write will overwrite all those values. This
data-type captures the semantics of the Amazon shopping cart
\cite{app:rep:optim:1606}, and the usual operation of Riak (when not using
CRDT data-types). Initial implementations of these registers tagged each
value with a full version vector \cite{syn:rep:sh143}; here we introduce an
optimized implementation that tags each value with a single dot, by using a
$\DotFun{V}$ as dot store. In Figure \ref{fig:MVReg} we can see that the write
delta mutator returns a causal context with all dots in the store, so that they are
removed upon join, together with a single mapping from a new dot to the value
written; as usual, the new dot is also put in the context.  A clear operation
simply removes current dots, leaving the register in the initial empty state.
Reading simply returns all values mapped in the store.

\begin{figure}[t]
\begin{eqnarray*}
  \afpar{MVRegister}{V} &=& \Causal{\DotFun{V}} \\
  \af{write}_i^\delta(v,(m,c)) &=& (\{d \mapsto v\}, \{d\} \union \dom m)
  \quad \kw{where} d = \af{next}_i(c) \\
  \af{clear}_i^\delta((m,c)) &=& (\{\}, \dom m) \\
  \af{read}((m,c)) & = & \ran m
\end{eqnarray*}
%
\caption{\dsrdt Multi-value register, replica $i$.}
\label{fig:MVReg}
\end{figure}

\subsubsection{Add-Wins Set} In an \emph{add-wins set} removals do not
affect elements that have been concurrently added. In this sense, under
concurrent updates, an add will win over a remove of the same element. The
implementation, in Figure \ref{fig:AWSet}, uses a map from elements to sets of
dots as dot store. This data-type can be seen as a map from elements to
enable-wins flags, but with a single common causal context, and keeping only
elements mapped to an enabled flag.

\begin{figure}[t]
\begin{eqnarray*}
  \afpar{AWSet}{E} &=& \Causal{\DotMap{E, \DotSet}} \\
  \af{add}_i^\delta(e,(m,c)) &=& (\{e \mapsto d\}, d \union m(e))
  \quad \kw{where} d = \{\af{next}_i(c)\} \\
  \af{remove}_i^\delta(e, (m,c)) &=& (\{\}, m(e)) \\
  \af{clear}_i^\delta((m,c)) &=& (\{\}, \af{dots}(m)) \\
  \af{elements}((m,c)) & = & \dom m
\end{eqnarray*}
\caption{\dsrdt Add-wins set, replica $i$.}
\label{fig:AWSet}
\end{figure}

When an element is added, all dots in the corresponding entry will be replaced
by a singleton set containing a new dot.  If a $\DotSet$ for some element were
to become empty, such as when removing the element, the join for $\DotMap{E,
\DotSet}$ will remove the entry from the resulting map. Concurrently created
dots are preserved when joining.  The $\af{clear}$ delta mutator will put all dots
from the dot store in the causal context, to be removed when joining.  As only
non-empty entries are kept in the map, the set of elements corresponds to the
map domain.

\subsubsection{Remove-Wins Set} Under concurrent adds and removes of the same
element, a \emph{remove-wins set} will make removes win.
To obtain this behaviour, the implementation in Figure \ref{fig:RWSet} uses
a map from elements to a nested map from booleans to sets of dots.  For both
adding and removing of a given entry, the corresponding nested map is cleared
(by the delta mutator inserting all corresponding dots into the causal
context), and a new mapping from either $\true$ or $\false$, respectively, to
a singleton new dot is added.

\begin{figure}[t]
\begin{eqnarray*}
  \afpar{RWSet}{E} &=& \Causal{\DotMap{E, \DotMap{\bool, \DotSet}}} \\
  \af{add}_i^\delta(e,(m,c)) &=& (\{e \mapsto \{\true \mapsto d\}\}, d \union \af{dots}(m(e)))
  \quad \kw{where} d = \{\af{next}_i(c)\} \\
  \af{remove}_i^\delta(e,(m,c)) &=& (\{e \mapsto \{\false \mapsto d\}\}, d \union \af{dots}(m(e)))
  \quad \kw{where} d = \{\af{next}_i(c)\} \\
  \af{clear}_i^\delta((m,c)) &=& (\{\}, \af{dots}(m)) \\
  \af{elements}((m,c)) & = & \{ e \in \dom m | \false \not\in \dom m(e) \}
\end{eqnarray*}
\caption{\dsrdt Remove-wins set, replica $i$.}
\label{fig:RWSet}
\end{figure}

When joining replicas, the nested map will collect the union of the respective
sets in the corresponding entry (for dots not seen by the other causal
context). As before, only non-bottom entries are kept, for both outer map
(non-empty maps) and nested map (non-empty $\DotSet$s). Therefore, an element
is considered to be in the set if it belongs to the outer map domain, and the
corresponding nested map does not contain a $\false$ entry;
thus, concurrent removes will win over adds.

\subsubsection{A Map Embedding Causal \dsrdt{}s.}
\label{sec:map}

Maps are important composition tools for the construction of complex CRDTs.
Although grow-only maps are simple to conceive and have been used in early
state based designs \cite{app:1639}, the creation of a map that allows removal
of entries and supports recursive composition is not trivial. Riak 2.0
introduced a map design that provides a clear observed-remove semantics: a
remove can be seen as an ``undo'' of all operations leading to the embedded
value, putting it in the bottom state, but remembering those operations, to
undo them in other replicas which observe it by a join.
Key to the design is to enable removal of keys to affect (and remember) the dots
in the associated nested CRDT, to allow joining with replicas that have
concurrently evolved from the before-removal point, or to ensure that
re-creating entries previously removed does not introduce anomalies.

In order to obtain the desired semantics it is not possible to simply map keys
to causal CRDTs having their own causal contexts. Doing so would introduce
anomalies when recreating keys, since old versions of the mappings
in other replicas could be considered more recent than newer mappings, since
the causal contexts of the re-created entries would start again at their
bottom state. The solution is to have a common causal context to the whole
map, to be used for all nested components, and never reset that single context.

For an arbitrary set of keys $K$ and a causal \dsrdt $\Causal{V}$ that we want
to embed (including, recursively, the map we are defining), the desired map
can be achieved through $\Causal{\DotMap{K,V}}$, where a single causal context
is shared by all keys and corresponding nested CRDTs, as presented in
Figure~\ref{fig:ORMap}.
This map can embed any causal CRDT as values. For instance we can define
a map of type $\afpar{ORMap}{S, \afpar{AWSet}{E}}$, mapping strings $S$ to
add-wins sets of elements $E$; or define a more complex recursive structure
that uses a map within a map
$\afpar{ORMap}{\nat, \afpar{ORMap}{S, \afpar{MVReg}{E}}}$.

\begin{figure}[t]
\begin{eqnarray*}
  \afpar{ORMap}{K, \Causal{V}} &=& \Causal{\DotMap{K,V}} \\
  \af{apply}_i^\delta(o_i^\delta, k, (m,c)) & = & (\{ k \mapsto v \}, c')
  \quad \kw{where} (v,c') = o_i^\delta((m(k),c)) \\
  \af{remove}_i^\delta(k,(m,c)) & = & (\{\}, \af{dots}(m(k))) \\
  \af{clear}_i^\delta((m,c)) & = & (\{\}, \af{dots}(m)) \\
\end{eqnarray*}
\caption{\dsrdt Map embedding Causal \dsrdt{}s, with observed removes, replica $i$.}
\label{fig:ORMap}
\end{figure}

The map does not support a specific operation to add new entries: it starts as
an empty map, which corresponds to any key implicitly mapped to bottom; then,
any operation from the embedded type can be applied, through a higher-order
$\af{apply}$, which takes a delta mutator $o_i^\delta$ to be applied, the key
$k$, and the map $(m,c)$. This mutator fetches the value at key $k$ from $m$,
pairs it with the shared causal context $c$, obtaining a value from the
embedded type, and invokes the operation over the pair; from the resulting
pair, it extracts the value to create a new mapping for that key, which it
pairs with the resulting causal context.  Removing a key will recursively
remove the dots in the corresponding embedded value, while the clear operation
will remove all dots from the store. This simplicity was achieved by
encapsulating most complexity in the join (and also the $\af{dots}$ function)
of the embedded type.


\section{Related Work}
\label{sec:related}
\subsection{Eventually convergent data types.}

The design of replicated systems that are always available and eventually
converge can be traced back to historical designs in
\cite{app:rep:optim:1501,db:rep:optim:1454}, among others.  More recently, replicated data types that
always eventually converge, both by reliably broadcasting operations (called
operation-based) or gossiping and merging states (called state-based), have
been formalized as
CRDTs~\cite{alg:rep:sh132,app:1639,rep:syn:sh138,syn:rep:sh143}. These are also
closely related to Bloom$^L$ \cite{conway2012logic} and Cloud
Types~\cite{burckhardt2012cloud}. 
State join-semilattices were used for deterministic parallel programming in
LVars~\cite{kuper2013lvars}, where variables progress in the lattice order by
joining other values, and are only accessible by special threshold reads. 


\subsection{Message size.}

A key feature of \dsrdt is message size reduction and coalescing, using
small-sized deltas. The general old idea of using differences between things,
called ``deltas'' in many contexts, can lead to many designs, depending on how
exactly a delta is defined.  The state-based deltas introduced for
Computational CRDTs ~\cite{navalho2013incremental} require an extra
delta-specific merge (in addition to the standard join) which does not ensure
idempotence.  In~\cite{deltaICDCS13}, an improved synchronization method for
non-optimized OR-set CRDT~\cite{rep:syn:sh138} is presented, where delta
information is propagated; in that paper deltas are a collection of items
(related to update events between synchronizations), manipulated and merged
through a protocol, as opposed to normal states in the semilattice. No generic
framework is defined (that could encompass other data types) and the protocol
requires several communication steps to compute the information to exchange.
Operation-based CRDTs~\cite{rep:syn:sh138,syn:rep:sh143,BAS2014} also support
small message sizes, and in particular, \emph{pure} flavors~\cite{BAS2014} that
restrict messages to the operation name, and possible arguments.  Though pure
operation-based CRDTs allow for compact states and are very fast at the source
(since operations are broadcast without consulting the local state), the model
requires more systems guarantees than \dsrdt do, e.g., exactly-once reliable
delivery and membership information, and impose more complex integration of new
replicas.  The work in \cite{export:211340} shows a different trade-off among
state deltas and pure operations, by tagging operations and creating a globally
stable log of operations while allowing local transient logs to preserve
availability.  While having other advantages, the creation of this global log
requires more coordination than our gossip approach for causally consistent
delta dissemination, and can stall dissemination.


\subsection{Encoding causal histories.}

State-based CRDT are always designed to be causally
consistent~\cite{app:1639,syn:rep:sh143}. Optimized implementations of sets,
maps, and multi-value registers can build on this assumption to keep the
meta-data small~\cite{DBLP:conf/popl/BurckhardtGYZ14}.  In \dsrdt, however,
deltas and delta-groups are normally not causally consistent, and thus the
design of \emph{join}, the meta-data state, as well as the anti-entropy
algorithm used must ensure this. Without causal consistency, the causal context
in \dsrdt can not always be summarized with version vectors, and consequently,
techniques that allow for gaps are often used.  A well known mechanism that
allows for encoding of gaps is found in Concise Version
Vectors~\cite{malkhi2007concise}. Interval Version Vectors~\cite{MGS14}, later
on, introduced an encoding that optimizes sequences and allows gaps, while
preserving efficiency when gaps are absent.



\section{Conclusion}
\label{sec:con}


We introduced the new concept of \dsrdt{}s and devised \emph{delta-mutators} over
state-based datatypes which can detach the changes
that an operation induces on the state. This brings a significant
performance gain as it allows only shipping small states, i.e., \emph{deltas},
instead of the entire state. The significant property in \dsrdt is that it
preserves the crucial properties (idempotence, associativity and
commutativity) of standard state-based CRDT.
 In addition, we have shown how \dsrdt can achieve 
causal consistency; and we presented an anti-entropy algorithm that allows replacing classical
state-based CRDTs by more efficient ones, while preserving their properties.
As an application of our approach we designed several novel \dsrdt specifications, including a general framework for causal CRDTs and composition in maps. 

Our approach is more relaxed than classical state-based CRDTs, and thus, can
replace them without losing their power since \dsrdt allows shipping delta-states
as well as the entire state. 
Another interesting observation is that \dsrdt can mimic the behavior of operation-based CRDTs, by
shipping individual deltas on the fly but with weaker guarantees from
the dissemination layer. 

\bibliographystyle{splncs}
\bibliography{predef,ref.bib,bib,shapiro-bib,local}


\end{document}